\documentclass[12pt,eqno]{article}
\usepackage{amsmath}
\numberwithin{equation}{section}

\newcommand{\alp}{\alpha}
\newcommand{\bt}{\beta}
\newcommand{\gm}{\gamma}

\newcommand{\dlt}{\delta}

\newcommand{\ep}{\epsilon}
\newcommand{\tht}{\theta}
\newcommand{\btht}{\bar{\tht}}
\newcommand{\kp}{\kappa}
\newcommand{\lmd}{\lambda}
\newcommand{\Lmd}{\Lambda}
\newcommand{\sgm}{\sigma}
\newcommand{\Sgm}{\Sigma}
\newcommand{\vph}{\varphi}
\newcommand{\omg}{\omega}
\newcommand{\Omg}{\Omega}

\newcommand{\ztR}{\zeta_{\rm R}}

\newcommand{\be}{\begin{equation}}
\newcommand{\ee}{\end{equation}}
\newcommand{\bea}{\begin{eqnarray}}
\newcommand{\eea}{\end{eqnarray}}
\newcommand{\eql}{\!\!\!&=\!\!\!&}

\newcommand{\defa}{\!\!\!&\equiv\!\!\!&}

\newcommand{\mtrx}[4]{\brkt{\begin{array}{cc}#1&#2\\#3&#4\end{array}}}

\newcommand{\vct}[2]{\brkt{\begin{array}{c}#1\\#2\end{array}}}

\newcommand{\tl}[1]{\tilde{#1}}
\newcommand{\bdm}[1]{{\mbox{\boldmath $#1$}}}

\newcommand{\diag}{{\rm diag}}
\newcommand{\der}{\partial}
\newcommand{\dr}{\!\!d}
\newcommand{\hc}{{\rm h.c.}}
\newcommand{\ie}{{\it i.e.}}
\newcommand{\id}{\mbox{\boldmath $1$}}
\newcommand{\kdlt}[2]{\delta_{#1}^{\;\;#2}}

\newcommand{\vev}[1]{\langle #1 \rangle}
\newcommand{\Lvev}[1]{\left\langle #1 \right\rangle}
\newcommand{\brkt}[1]{\left( #1 \right)}
\newcommand{\brc}[1]{\left\{ #1 \right\}}
\newcommand{\sbk}[1]{\left[ #1 \right]}
\newcommand{\abs}[1]{\left| #1 \right|}

\renewcommand{\Re}{{\rm Re}}
\renewcommand{\Im}{{\rm Im}}

\newcommand{\cA}{{\cal A}}
\newcommand{\cD}{{\cal D}}
\newcommand{\cF}{{\cal F}}
\newcommand{\cH}{{\cal H}}
\newcommand{\cL}{{\cal L}}
\newcommand{\cM}{{\cal M}}
\newcommand{\cN}{{\cal N}}
\newcommand{\cO}{{\cal O}}
\newcommand{\cP}{{\cal P}}
\newcommand{\cR}{{\cal R}}

\newcommand{\cV}{{\cal V}}
\newcommand{\cW}{{\cal W}}

\newcommand{\gomg}[2]{\omega_{#1}^{\;\;#2}}
\renewcommand{\ge}[2]{e_{#1}^{\;\;#2}}

\newcommand{\dQ}{\dlt}
\newcommand{\udl}[1]{\underline{#1}}

\newcommand{\dmx}{d_{\alp}^{\;\;\bt}}
\newcommand{\nV}{n_{\rm V}}
\newcommand{\nH}{n_{\rm H}}
\newcommand{\fG}{f_G}
\newcommand{\fh}{f_{\rm h}}
\newcommand{\DV}{D_{\rm V}}
\newcommand{\fS}{f_S}
\newcommand{\gey}{\vev{\ge{y}{4}}}
\newcommand{\SUu}{SU(2)_{\mbox{\scriptsize $\bdm{U}$}}}
\newcommand{\Usp}{Usp(2,2\nH)}
\newcommand{\lrder}{\stackrel{\leftrightarrow}{\partial}}
\newcommand{\tlg}{\mbox{\scriptsize $\bdm{\tl{g}}$}}

\newcommand{\NP}[1]{{\it Nucl.~Phys.}~{\bf #1}}
\newcommand{\PL}[1]{{\it Phys.~Lett.}~{\bf #1}}

\newcommand{\PR}[1]{{\it Phys.~Rev.}~{\bf #1}}
\newcommand{\PRL}[1]{{\it Phys.~Rev.~Lett.}~{\bf #1}}
\newcommand{\PTP}[1]{{\it Prog.~Theor.~Phys.}~{\bf #1}}

\newcommand{\JH}[1]{{\it JHEP}~{\bf #1}}

\begin{document}

\begin{titlepage}
\null
\begin{flushright}
 {\tt hep-th/0408224}\\
KAIST-TH 2004/11
\\
August, 2004
\end{flushright}

\vskip 2cm
\begin{center}
\baselineskip 0.8cm
{\LARGE \bf Superfield description of 5D supergravity \\
on general warped geometry}

\lineskip .75em
\vskip 2.5cm

\normalsize

{\large\bf Hiroyuki Abe}{\def\thefootnote{\fnsymbol{footnote}}
\footnote[1]{\it e-mail address:abe@muon.kaist.ac.kr}}
{\large\bf and Yutaka Sakamura}{\def\thefootnote{\fnsymbol{footnote}}
\footnote[2]{\it e-mail address:sakamura@muon.kaist.ac.kr}}

\vskip 1.5em

{\it Department of Physics, \\
Korea Advanced Institute of Science and Technology \\
Daejeon 305-701, Korea}

\vspace{18mm}

{\bf Abstract}\\[5mm]
{\parbox{13cm}{\hspace{5mm} \small
We provide a systematic and practical method 
of deriving 5D supergravity action 
described by 4D superfields on a {\it general warped geometry}, 
including a non-BPS background. 
Our method is based on the superconformal formulation of 5D supergravity, 
but is easy to handle thanks to the superfield formalism. 
We identify the radion superfield in the language of 5D superconformal 
gravity, and clarify its appearance in the action. 
We also discuss SUSY breaking effects induced by a deformed geometry 
due to the backreaction of the radius stabilizer. 
}}

\end{center}

\end{titlepage}

\clearpage

\section{Introduction}
In recent years, the brane-world scenario~\cite{ADD,RS} has attracted 
much attention as a candidate for the new physics beyond the standard model 
and has been investigated in various frameworks. 
The idea that our four-dimensional (4D) world is embedded into 
a higher dimensional spacetime was introduced\footnote{
In the context of field theory, such an idea was originally studied 
in Ref.~\cite{rubakov}. } 
by Ho\v{r}ava and Witten in their investigation 
of the strongly-coupled heterotic string theory~\cite{HW}. 
Five-dimensional supergravity (5D SUGRA) 
on an orbifold~$S^1/Z_2$ appears as a low-energy effective theory 
after the reduction of their theory to five dimensions by compactifying 
on a Calabi-Yau 3-fold~\cite{LOSW}. 
In addition, 5D SUGRA has been investigated actively 
as a simplest stage of the supersymmetric (SUSY) brane-world scenario. 

When we construct brane-world models, a size of the extra dimensions 
has to be stabilized. 
One of the main stabilization mechanisms is proposed in Ref.~\cite{GW}, 
and similar mechanisms have also been studied~\cite{dewolfe,EMS,MO}. 
These mechanisms involve a bulk scalar field that has a nontrivial field 
configuration. 
Generically, such a scalar configuration does not saturate 
Bogomol'nyi-Prasad-Sommerfield (BPS) bound~\cite{BPS}, 
and breaks SUSY~\cite{EMS}. 
At the same time, the background geometry receives the backreaction 
of the scalar configuration, which has not been taken into account 
in most works.  
Thus, SUSY breaking effects can be mediated to our visible sector 
through the deformation of the spacetime geometry 
even if the visible sector is decoupled from the stabilization sector 
where SUSY breaking occurs. 
One of the authors discussed such SUSY breaking effects 
in a toy model where the bulk spacetime is four dimensions 
and the effective theory is three-dimensional~\cite{geoSB4D}. 
In that work, a systematic method is proposed 
to describe a 4D SUGRA action in terms of 3D superfields 
on a {\it general warped geometry}, including a non-BPS background. 
In this paper, we will apply this method to 5D SUGRA 
and derive 5D SUGRA action described by 4D superfields 
on the general warped geometry. 

For concreteness of the discussion, we will assume that 
the radius of the extra dimension is stabilized by a scalar configuration 
of a hypermultiplet, which is decoupled from the visible sector. 
In this case, the spacetime geometry is determined by the nontrivial vacuum 
configuration of the radius stabilizer in the hidden sector 
besides the cosmological constant through the Einstein equation. 
We will also assume that the extra dimension is compactified on 
an orbifold~$S^1/Z_2$. 
In this paper, a characteristic scale of the hidden sector~$\Lmd_{\rm hid}$ 
is supposed to be much smaller than the 5D Planck mass~$M_5$. 
For example, $\Lmd_{\rm hid}$ is supposed to be an intermediate scale. 
Thus, interactions with the gravitational fields can be neglected. 
Especially, in the limit of $\Lmd_{\rm hid}/M_5\to 0$, 
the deformation of the geometry vanishes and the spacetime becomes 
flat or a slice of anti-de Sitter (AdS) space. 

The superconformal formulation of 5D SUGRA is a systematic and elegant way 
of describing the bulk-boundary system~\cite{KO3,KO1,KO2,Ohashi}, 
but it involves very complicated and tedious calculations 
to derive the action. 
Our method is based on the superconformal formulation, 
but is much easier to handle thanks to its 4D superfield description. 
Thus, our method is useful as a practical method of deriving 
the 4D effective action expressed by superfields. 

The radion superfield plays an important role in transmission of 
SUSY breaking effects to our visible sector~\cite{marti,chacko}. 
Although there are some papers identifying the radion chiral multiplet 
with fields in the off-shell 5D SUGRA~\cite{quiros}, 
no one has constructed the radion {\it superfield} 
from the off-shell SUGRA directly. 
Using our method, we can identify the radion superfield 
and see its appearance in the action directly 
from the off-shell SUGRA. 

The paper is organized as follows. 
In the next section, we will provide an invariant action 
expressed by 4D superfields. 
In Sect.~\ref{gauge_fix}, we will explain the gauge fixing 
of the extraneous superconformal symmetry, and specify 
the physical superfields. 
In Sect.~\ref{EFT}, we will derive the action 
after the gauge fixing, and identify the radion superfield. 
SUSY breaking induced by the geometry is also discussed. 
Sect.~\ref{summary} is devoted to the summary. 
The notation we use in this paper is listed in Appendix~\ref{notation}, 
and brief comments on the invariant action is provided 
in Appendix~\ref{inv_act}.

\section{Off-shell formulation}
In this section, we will briefly explain the field content in 
the superconformal formulation of 5D SUGRA.  
Then, we will construct 4D superfields\footnote{ 
In this paper, we will use the word ``4D superfield'' 
for an $\cN=1$ superfield that is allowed to have $y$-dependence.  
} and express an invariant action in terms of them. 
Basically, we will follow the notation of Ref.~\cite{KO2}. 
Throughout this paper, we will use $\mu,\nu,\cdots=0,1,2,3,4$ 
for the 5D world vector indices, and $m,n,\cdots=0,1,2,3$ 
for the 4D indices. 
The coordinate of an extra dimension is denoted as $y\equiv x^4$. 
The corresponding local Lorentz indices are denoted by underbarred indices. 

\subsection{Field contents}
The 5D Weyl multiplet consists of the following fields. 
\be
 \ge{\mu}{\udl{\nu}}, \;\;\;
 \psi_\mu^i, \;\;\;
 V^{ij}_\mu, \;\;\;
 b_\mu, \;\;\;
 v_{\udl{\mu}\udl{\nu}}, \;\;\;
 \chi^i, \;\;\;
 D, 
\ee
where $i,j=1,2$ are indices of $\SUu$, which is related to 
$SU(2)_R$ after the superconformal gauge fixing. 
These are the f\"{u}nfbein, the gravitino, the gauge bosons for 
$\SUu$ and the dilatation, 
a real antisymmetric tensor, an $\SUu$ Majorana spinor, and a real scalar, 
respectively. 
Among these, only $\ge{\mu}{\udl{\nu}}$ and $\psi_\mu^i$ are dynamical. 
$b_\mu$ is eliminated by the gauge fixing condition, 
and the remaining ones are auxiliary fields. 
Especially, $\chi^i$ and $D$ appear in the action in the form 
of Lagrange multipliers \cite{KO1}. 

In this paper, we will introduce vector multiplets~$\cV^I$ 
($I=1,2,\cdots,\nV$)
and hypermultiplets~$\cH^{\hat{\alp}}$ ($\hat{\alp}=1,2,\cdots,\nH$) 
as matter multiplets. 
In addition to these, there is a vector multiplet~$\cV^0$ that includes 
the graviphoton,\footnote{In Ref.~\cite{KO2}, this is called 
the central charge vector multiplet. } 
and a compensator hypermultiplet~$\cH^0$.\footnote{
We will assume that there is only one compensator 
hypermultiplet. An extension to the multi-compensator case 
is straightforward. }
Thus, there are $\nV+1$ vector multiplets~$\cV^I$ ($I=0,1,2,\cdots,\nV$), 
and $\nH+1$ hypermultiplets~$\cH^{\hat{\alp}}$ 
($\hat{\alp}=0,1,2,\cdots,\nH$).  

A vector multiplet~$\cV^I$ consists of 
\be
 M^I,\;\;\; W_\mu^I,\;\;\; \Omg^{Ii}, \;\;\; Y^{Iij}, 
\ee
which are a gauge scalar, a gauge field, a gaugino 
and an auxiliary field, respectively. 

The hypermultiplets consist of 
complex scalars~$\cA_i^\alp$, spinors~$\zeta^\alp$ 
and auxiliary fields~$\cF_i^\alp$. 
They carry a $\Usp$ index~$\alp$ ($\alp=1,2,\cdots,2\nH+2$) on which 
the gauge group~$G$ can act. 
These are split into $\nH+1$ hypermultiplets as 
\be 
 \cH^{\hat{\alp}}=(\cA_i^{2\hat{\alp}+1},\cA_i^{2\hat{\alp}+2}, 
 \zeta^{2\hat{\alp}+1},\zeta^{2\hat{\alp}+2}, 
 \cF_i^{2\hat{\alp}+1}, \cF_i^{2\hat{\alp}+2}). 
\ee
As mentioned above, $\cH^{\hat{\alp}=0}$ is used as a compensator multiplet.  
$\cH^{\hat{\alp}=1}$ is supposed to be a stabilizer multiplet 
whose scalar components have a nontrivial vacuum configuration 
that is relevant to the radius stabilization. 
The remaining hypermultiplets are matter fields in the visible sector. 

The hyperscalars~$\cA_i^\alp$ satisfy the hermiticity condition
\be
 \cA^i_\alp\equiv \ep^{ij}\cA_j^\bt\rho_{\bt\alp}=-(\cA^\alp_i)^*, 
 \label{hmt_cond}
\ee
where $\rho_{\alp\bt}$ is an antisymmetric tensor. 
(See Appendix~\ref{notation}.)
Thus, we can choose $\cA_{i=2}^\alp$ ($\alp=1,2,\cdots,2\nH+2$) 
as independent fields. 
The auxiliary fields~$\cF_i^\alp$ also satisfy the same type of condition. 

The gravitino~$\psi_\mu^i$ and the gauginos~$\Omg^{Ii}$ are 
$\SUu$ Majorana spinors, and the hyperinos~$\zeta^\alp$ are 
$\Usp$ Majorana spinors. 
Due to the corresponding Majorana conditions, we can choose 
only the right-handed components of these spinors as independent fields. 
Thus, we will use a 2-component spinor notation defined by Eq.(\ref{def_RH}) 
in the following sections. 
The relation to the 4-component spinor notation is referred 
in Appendix~\ref{notation}. 

$V_\mu^{ij}$ and $Y^{Iij}$ are triplets of $\SUu$. 
In the following, we will use the `isovector' notation 
for these triplets, \ie, 
\be
 V_\mu^{ik}\ep_{kj} \equiv i\sum_{r=1}^3V_\mu^{(r)}(\sgm_r)^i_{\;\;j}, 
 \;\;\;\;\;
 Y^{Iik}\ep_{kj} \equiv i\sum_{r=1}^3Y^{I(r)}(\sgm_r)^i_{\;\;j}, 
\ee
where $\ep_{ij}$ is an antisymmetric tensor ($\ep_{12}=1$), 
and $\sgm_r$ ($r=1,2,3$) are the Pauli matrices. 

The orbifold parity of each field is listed in Table~\ref{Z2_parity}. 
Here, we will assume that all the physical vector multiplets have 
massless 4D vector fields in the Kaluza-Klein decomposition, 
while the graviphoton has an odd parity and no massless mode. 
Namely, $\Pi_{I=0}=-1$ and $\Pi_{I\neq 0}=+1$. 
For the hypermultiplets, we can always redefine the fields 
by using a degree of freedom for $\Usp$ 
so that their parity eigenvalues are those listed in the table. 
\begin{table}[t]
\begin{center}
\begin{tabular}{c|c} \hline\hline
\multicolumn{2}{c}{Weyl multiplet} \\ \hline
\rule[-2mm]{0mm}{7mm}$\Pi=+1$ & $\ge{m}{\udl{n}},\ge{y}{4},\psi_{m{\rm R}}^1,
\psi_{y{\rm R}}^2,b_m, V_m^{(3)}, V_y^{(1,2)}, v_{4\udl{m}}, \chi_{\rm R}^1, D$
 \\ \hline
\rule[-2mm]{0mm}{7mm}$\Pi=-1$ & $\ge{m}{4}, \ge{y}{\udl{n}}, 
\psi_{m{\rm R}}^2, \psi_{y{\rm R}}^1, 
b_y, V_y^{(3)}, V_m^{(1,2)}, v_{\udl{m}\udl{n}}, \chi_{\rm R}^2$ \\
\hline\hline
\multicolumn{2}{c}{Vector multiplet} \\ \hline
\rule[-2mm]{0mm}{7mm}$\Pi_I$ &  
$W_m^I,\;\;\; \Omg_{\rm R}^{I1},\;\;\; Y^{I(3)}$
\\ \hline
\rule[-2mm]{0mm}{7mm}$-\Pi_I$ & 
$M^I,\;\;\; W_y^I,\;\;\; \Omg_{\rm R}^{I2},\;\;\; Y^{I(1,2)}$ 
\\ \hline\hline
\multicolumn{2}{c}{Hypermultiplet} \\ \hline
\rule[-2mm]{0mm}{7mm}$\Pi=+1$ &  
$\cA_{i=2}^{2\hat{\alp}+2},\;\;\; \cF_{i=1}^{2\hat{\alp}+2},\;\;\; 
\zeta_{\rm R}^{2\hat{\alp}+2}$ \\ \hline
\rule[-2mm]{0mm}{7mm}$\Pi=-1$ & 
$\cA_{i=2}^{2\hat{\alp}+1},\;\;\; \cF_{i=1}^{2\hat{\alp}+1},\;\;\; 
\zeta_{\rm R}^{2\hat{\alp}+1}$ \\ \hline
\end{tabular}
\end{center}
\caption{Orbifold parity eigenvalues}
\label{Z2_parity}
\end{table}

\subsection{Superfields and invariant action}
We will take the warped metric ansatz for the background geometry, 
\be
 ds^2=g_{\mu\nu}dx^\mu dx^\nu=e^{2\sgm(y)}\eta_{\udl{m}\udl{n}}dx^mdx^n
 -\gey^2 dy^2, 
\ee
where $\gey$ is a vacuum expectation value (VEV) of $\ge{y}{4}$. 

Since we are not interested in the gravitational interactions, 
which are suppressed by the large Planck mass~$M_5$, 
we will freeze the gravitational multiplet to its background value.\footnote{
The replacement of $v_{\udl{\mu}\udl{\nu}}$ with its background value 
causes short of some terms in the invariant action, 
but this is not a problem for our purpose. (See Appendix~\ref{inv_act}.) } 
\bea
 \vev{\ge{m}{\udl{n}}} \eql e^\sgm\kdlt{m}{n}, \;\;\;\;\; 
 \gey=\mbox{constant},  \nonumber\\
 \vev{\psi_\mu^i} \eql 0, \nonumber\\
 \vev{V_m^{(r)}} \eql 0, \;\;\; (r=1,2,3) \nonumber\\
 \vev{v_{\udl{\mu}\udl{\nu}}} \eql 0, 
 \label{grv_bgd}
\eea
where $\vev{\cdots}$ denotes a VEV of the argument. 
Note that $\psi_\mu^i$, $V_m^{(r)}$ and $v_{\udl{\mu}\udl{\nu}}$ cannot have 
non-zero VEVs because of the unbroken 4D Poincar\'{e} invariance. 
On the other hand, the extra-components of the $\SUu$ gauge 
fields~$V_y^{(r)}$ can have non-zero VEVs. 
To simplify the discussion, we will ignore the dependence on 
$\vev{V_y^{(3)}}$. 
We will comment on this point in Sect.~\ref{SSSB}. 
In the following, we will also drop the dependences on $b_\mu$, $\chi^i$ 
and $D$ because $b_\mu$ will be set to zero by the gauge fixing condition 
and $\chi^i$ and $D$ only play Lagrange multipliers as mentioned 
in the previous subsection. 
We can always rescale the coordinate~$y$ so that $\gey=1$, 
but here we will leave $\gey$ as an arbitrary constant 
in order to make clear the correspondence to the 4D superconformal multiplets 
in Ref.~\cite{KO2}. 

In the case of $\vev{V_y^{(r)}}=0$, we can construct 
4D superfields from 5D vector and hyper multiplets, 
and express an invariant action on the above gravitational background 
in terms of them by using a method proposed in Ref.~\cite{geoSB4D}. 
By introducing the spurion superfield\footnote{
This superfield corresponds to the 4D general multiplet~$\bdm{W}_y$ 
in Ref.~\cite{KO2}. } 
\be
 V_T=\gey+i\tht^2e^\sgm\vev{V_y^{(1)}+iV_y^{(2)}}
  -i\btht^2e^\sgm\vev{V_y^{(1)}-iV_y^{(2)}}, 
 \label{def_VT}
\ee
and modifying some auxiliary components of the above constructed superfields, 
we can incorporate $\vev{V_y^{(1)}}$ and $\vev{V_y^{(2)}}$ 
into the invariant action. 
Each superfield is defined as follows. 

From the vector multiplets, we will define the following 4D vector and 
chiral superfields. 
\bea
 V^I \defa \tht\sgm^{\udl{m}}\btht W^I_m+i\tht^2\btht\bar{\lmd}^I
 -i\btht^2\tht\lmd^I+\frac{1}{2}\tht^2\btht^2D^I, \nonumber\\
 \Phi^I_S \defa \vph_S^I-\tht\chi_S^I-\tht^2\cF_S^I, 
\eea
where\footnote{The minus signs of the coefficients of $\tht$ and $\tht^2$ 
in the chiral superfields are necessary for matching to 
the notation of Refs.~\cite{KO1,KO2,Ohashi}. }
\bea
 \lmd^I \defa 2e^{\frac{3}{2}\sgm}\Omg^{I1}_{\rm R}, \nonumber\\
 D^I \defa -e^{2\sgm}\brc{\gey^{-1}\der_yM^I-2Y^{I(3)}
 +\gey^{-1}\dot{\sgm}M^I}, \nonumber\\
 \vph_S^I \defa \frac{1}{2}\brkt{W_y^I+i\gey M^I}, \nonumber\\
 \chi_S^I \defa -2e^{\frac{\sgm}{2}}\gey\Omg^{I2}_{\rm R}, \nonumber\\
 \cF_S^I \defa e^\sgm\brc{\vev{V_y^{(1)}+iV_y^{(2)}}M^I
  -i\gey\brkt{Y^{I(1)}+iY^{I(2)}}}. 
 \label{sf_comp1}
\eea
The dot denotes derivative in terms of $y$. 
Here, we have chosen the Wess-Zumino gauge. 
For simplicity, we will consider only abelian gauge groups 
in this paper. 
Under the gauge transformation, the above superfields transform as 
\bea
 V^I &\to& V^I+\Lmd^I+\bar{\Lmd}^I, \nonumber\\
 \Phi_S^I &\to& \Phi_S^I+i\der_y\Lmd^I, \label{sg_trf}
\eea
where $\Lmd^I$ are arbitrary chiral superfields. 
Thus, the gauge invariant quantities are 
\bea
 V_S^I \defa \brkt{-\der_yV^I-i\Phi_S^I+i\bar{\Phi}_S^I}/V_T, \nonumber\\
 \cW^I_\alp \defa -\frac{1}{4}\bar{D}^2D_\alp V^I. 
\eea
Here, $V_T$ in the definition of $V_S^I$ is not necessary 
for the invariance under the transformation~(\ref{sg_trf}), 
but is necessary for covariantization of $y$-derivative 
in terms of $\SUu$. 

From the hypermultiplets, we will define the following chiral superfields. 
\be
 \Phi^\alp \equiv \vph^\alp-\tht\chi^\alp-\tht^2\cF^\alp, 
\ee
where
\bea
 \vph^\alp \defa \cA^\alp_2, \nonumber\\
 \chi^\alp \defa -2ie^{\frac{\sgm}{2}}\ztR^\alp,  \nonumber\\
 \cF^\alp \defa e^\sgm\gey^{-1}\left\{\der_y\cA_1^\alp
  +i\vev{V_y^{(1)}+iV_y^{(2)}}\cA_2^\alp
  +i\brkt{\gey+\frac{iW_y^0}{M^0}}\tl{\cF}_1^\alp \right.\nonumber\\
  &&\hspace{20mm}\left.-\brkt{W_y^I-i\gey M^I}(gt_I)^\alp_{\;\;\bt}\cA_1^\bt
  +\frac{3}{2}\dot{\sgm}\cA_1^\alp\right\}. 
 \label{sf_comp2}
\eea
In the last expression, $\tl{\cF}_1^\alp$ is defined as
\be
 \tl{\cF}_1^\alp \equiv \cF_1^\alp-M^0(gt_0)^\alp_{\;\;\bt}\cA_1^\bt. 
\ee
The generators of the gauge group~$t_I$ ($I=0,1,\cdots,\nV$) are defined 
as anti-hermitian. 

Using these superfields, the 5D invariant action can be written as follows. 
\bea
 S \eql \int\dr^5x\;\cL_{\rm vector}+\cL_{\rm hyper}, \nonumber\\
 \cL_{\rm vector} \eql \sbk{\int\dr^2\tht\;
 \frac{3C_{IJK}}{2}\brc{i\Phi_S^I\cW^J\cW^K
 +\frac{1}{12}\bar{D}^2\brkt{V^ID^\alp\der_yV^J-D^\alp V^I\der_yV^J}\cW^K_\alp}
 +\hc} \nonumber\\
 &&-e^{2\sgm}\int\dr^4\tht\;V_T C_{IJK}V_S^IV_S^JV_S^K, \nonumber\\ 
 \cL_{\rm hyper} \eql -2e^{2\sgm}\int\dr^4\tht\;
 V_T\dmx\bar{\Phi}^\bt\brkt{e^{-2igV^It_I}}^\alp_{\;\;\gm}\Phi^\gm 
 \nonumber\\
 && -e^{3\sgm}\sbk{\int\dr^2\tht\;
 \Phi^\alp\dmx\rho_{\bt\gm}\brkt{\der_y-2g\Phi_S^It_I}^\gm_{\;\;\dlt}\Phi^\dlt
 +\hc}, 
 \label{Sinv}
\eea
where $C_{IJK}$ is a real constant tensor which is completely symmetric 
for the indices, 
and $\dmx$ is a metric of the hyperscalar space and can be brought 
into the standard form~\cite{dewit} 
\be
 \dmx=\mtrx{\bdm{1}_2}{}{}{-\bdm{1}_{2\nH}}.  \label{def_dmx}
\ee

Eq.(\ref{Sinv}) is very similar to the action in Ref.~\cite{DGN}, 
except for the appearance of the spurion superfield~$V_T$ 
and the warp factor.\footnote{
The $V_T$-dependence in Eq.(\ref{Sinv}) is also taken into account 
in Ref.~\cite{CST}.} 
However, note that our action is SUGRA action 
{\it before the superconformal gauge fixing}, 
which is explained in the next section. 
The above action certainly reproduces the component action 
in Refs.~\cite{KO1,Ohashi}.\footnote{An invariant action constructed 
in Ref.~\cite{KO1} coincides with that of Ref.~\cite{Ohashi} 
on the gravitational background~(\ref{grv_bgd}), 
although $\bdm{S}$-gauge is already fixed from the beginning 
in Ref.~\cite{KO1}. }
(See Appendix~\ref{inv_act}.)
The first line of $\cL_{\rm vector}$ corresponds to the gauge kinetic terms 
and the supersymmetric Chern-Simons term. 
In fact, variation of these terms under 
the gauge transformation~(\ref{sg_trf}) does not vanish, 
but becomes total derivative. 
Thus, these terms cannot be expressed 
by only the gauge-invariant quantities~$V_S^I$ and $\cW^I$. 

In addition to the above action, we can also introduce the brane actions 
localized at the fixed boundaries of the orbifold. 
We will discuss them in Sect.~\ref{action_fml}.

\section{Gauge fixing and physical superfields} \label{gauge_fix}
\subsection{Gauge fixing conditions} \label{GFcond}
In order to obtain the Poincar\'{e} supergravity, 
we have to fix the extraneous superconformal symmetries, \ie, 
the dilatation~$\bdm{D}$, $\SUu$, the conformal supersymmetry~$\bdm{S}$ and 
the special conformal transformation~$\bdm{K}$. 
The gauge fixing condition for each symmetry is as follows~\cite{KO2,Ohashi}. 

The $\bdm{D}$-gauge is fixed by  
\bea
 \cN \defa C_{IJK}M^IM^JM^K=M_5^3, \label{Dfix1} \\
 \cA_i^\alp\dmx\cA_\bt^i \eql 2\brc{-\sum_{a=1}^2\abs{\cA_2^a}^2
 +\sum_{\udl{\alp}=3}^{2\nH+2}\abs{\cA_2^{\udl{\alp}}}^2}
 =-2M_5^3,  \label{Dfix2}
\eea
where $\cN$ is called a norm function, 
and $\SUu$ is fixed by the condition 
\be
 \cA_i^a \propto \dlt_i^a. \;\;\;\;\; (a=1,2)  \label{Ufix}
\ee

The $\bdm{S}$-gauge is fixed by 
\bea
 \cN_I\Omg^{Ii} \eql 0, \label{Sfix1} \\
 \cA_i^\alp\dmx\zeta_\bt \eql 0,  \label{Sfix2}
\eea
where $\cN_I\equiv\der\cN/\der M^I$. 

The $\bdm{K}$-gauge fixing condition is
\be
 b_\mu=0, \label{Kfix}
\ee
which is already taken into account in our case.

Here, note that the compensator scalars~$\cA_i^a$ ($a=1,2$) 
are charged under both $\SUu$ 
and an $SU(2)$ subgroup of $\Usp$ that rotates 
only the compensator part, which refer to as $SU(2)_C$. 
Thus, the $\SUu$ gauge fixing condition~(\ref{Ufix}) breaks 
$\SUu\times SU(2)_C$ up to an $SU(2)$ diagonal subgroup. 
This is the symmetry known as $SU(2)_R$ in the Poincar\'{e} supergravity. 

In the original superconformal formulation, the conditions~(\ref{Dfix1}) 
and (\ref{Dfix2}), or (\ref{Sfix1}) and (\ref{Sfix2}) are not independent 
because the equations of motion for $\chi^i$ and $D$ in the Weyl multiplet 
lead to the constraints 
\bea
 \cA_i^\alp\dmx\cA_\bt^i+2\cN \eql 0, \label{ctrt1} \\
 \cA_i^\alp\dmx\zeta_\bt+\cN_I\Omg^I_i \eql 0. \label{ctrt2}
\eea
In our case, however, we must impose these constraints by hand 
since we have dropped the dependences of $\chi^i$ and $D$. 

In the vector sector, we will consider the maximally symmetric case 
in the following. 
Namely, the norm function is 
\be
 \cN = (M^{I=0})^3-\frac{1}{2}M^{I=0}\sum_{J=1}^{\nV}(M^J)^2. 
 \label{norm_fct}
\ee
In this case, the gauge scalars~$M^I$ which satisfy 
the constraint~(\ref{Dfix1}) are parametrized 
by the physical scalar fields~$\phi^x$ ($x=1,\cdots,\nV$) as follows. 
\bea
 M^{I=0} \eql M_5\brc{\cosh\hat{\phi}^1\cosh\hat{\phi}^2
 \cdots\cosh\hat{\phi}^{\nV}}^{2/3} 
 = M_5+\frac{1}{3M_5^2}\sum_{x=1}^{\nV} (\phi^x)^2+\cO(M_5^{-5}), \nonumber\\
 M^{I=1} \eql \frac{\sqrt{2}M_5\sinh\hat{\phi}^1}
 {\brc{\cosh\hat{\phi}^1\cosh\hat{\phi}^2\cdots\cosh\hat{\phi}^{\nV}}^{1/3}}
 =\frac{\sqrt{2}}{M_5^{1/2}}\phi^1+\cO(M_5^{-7/2}), \nonumber\\
 M^{I=2} \eql \frac{\sqrt{2}M_5\cosh\hat{\phi}^1\sinh\hat{\phi}^2}
 {\brc{\cosh\hat{\phi}^1\cosh\hat{\phi}^2\cdots
  \cosh\hat{\phi}^{\nV}}^{1/3}}
 =\frac{\sqrt{2}}{M_5^{1/2}}\phi^2+\cO(M_5^{-7/2}), \nonumber\\
 &\vdots& \nonumber\\
 M^{I=\nV} \eql \frac{\sqrt{2}M_5\cosh\hat{\phi}^1
 \cosh\hat{\phi}^2\cdots\cosh\hat{\phi}^{\nV-1}\sinh\hat{\phi}^{\nV}}
 {\brc{\cosh\hat{\phi}^1\cosh\hat{\phi}^2\cdots
  \cosh\hat{\phi}^{\nV}}^{1/3}} 
 = \frac{\sqrt{2}}{M_5^{1/2}}\phi^{\nV}+\cO(M_5^{-7/2}),  \nonumber\\
 \label{Mtophi}
\eea
where $\hat{\phi}^x\equiv\phi^x/M_5^{3/2}$. 

In the language of 4D superfields in the previous section, 
the gauge fixing conditions~(\ref{Dfix2}-\ref{Sfix2}) can be written 
as follows. 
\bea
 \vph^{\alp=1} \eql 0, \nonumber\\
 \vph^{\alp=2} \eql \brkt{M_5^3
 +\sum_{\udl{\alp}=3}^{2\nH+2}\abs{\vph^{\udl{\alp}}}}^{1/2}
 =M_5^{3/2}+\frac{1}{2M_5^{3/2}}\sum_{\udl{\alp}=3}^{2\nH+2}
 \abs{\vph^{\udl{\alp}}}^2+\cO(M_5^{-9/2}). \nonumber\\
 \chi^{\alp=1} \eql -\frac{1}{M_5^{3/2}}\sum_{\udl{\bt},\udl{\gm}=3}^{2\nH+2}
  \rho_{\udl{\bt}\udl{\gm}}\vph^{\udl{\bt}}\chi^{\udl{\gm}}
  +\cO(M_5^{-9/2}), \nonumber\\
 \chi^{\alp=2} \eql \frac{1}{M_5^{3/2}}\sum_{\udl{\bt}=3}^{2\nH+2}
 \bar{\vph}^{\udl{\bt}}\chi^{\udl{\bt}}+\cO(M_5^{-9/2}), \nonumber\\
 \lmd^{I=0} \eql \frac{\sqrt{2}}{3M_5^{3/2}}\sum_{x=1}^{\nV}\phi^x\lmd^{I=x} 
 +\cO(M_5^{-9/2})  \nonumber\\
 \chi_S^{I=0} \eql \frac{\sqrt{2}}{3M_5^{3/2}}\sum_{x=1}^{\nV}
 \phi^x\chi_S^{I=x}+\cO(M_5^{-9/2}). 
\eea

\subsection{Physical gauge superfields}
Due to the gauge fixing conditions~(\ref{Dfix1}) and (\ref{Sfix1}), 
the physical gauge scalars and gauginos in 5D vector multiplets 
have only $\nV$ components, respectively.  
Thus, the $\nV+1$ vector multiplets can be expressed by $\nV$ physical vector 
and chiral superfields~$\tl{V}^x$ and $\tl{\Phi}_S^x$ ($x=1,\cdots,\nV$) 
besides the graviphoton~$B_\mu$ after the gauge fixing. 
A manifold~$\cM$ of the physical gauge scalars~$\phi^x$ is 
the very special manifold. 
Various geometrical quantities of $\cM$ 
are as follows~\cite{GST}. 
\bea
 h^I(\phi) \defa -\sqrt{\frac{2}{3}}M^I(\phi), \;\;\;\;\;
 h_I(\phi) \equiv -\frac{1}{\sqrt{6}}\cN_I, \;\;\;\;\;
 h^I_x(\phi) \equiv \frac{\der M^I}{\der\phi^x}, \nonumber\\
 a_{IJ} \defa -\frac{1}{2}\frac{\der^2}{\der M^I\der M^J}\ln\cN, \;\;\;\;\;
 a^{IJ} \equiv (a^{-1})^{IJ}, \nonumber\\
 g_{xy} \defa a_{IJ}h^I_x h^J_y, \;\;\;\;\;
 g^{xy} \equiv (g^{-1})^{xy},  \label{geo_qnt}
\eea
where $a_{IJ}$ is a metric of the ambient $\nV+1$ dimensional space, 
and $g_{xy}$ is an induced metric on $\cM$. 

From Eq.(\ref{Mtophi}), the physical gauge scalars can be expressed as 
\be
 \phi^x=\frac{M_5^{1/2}}{\sqrt{2}}M^{I=x}+\cO(M_5^{-5/2}). 
 \label{phitoM}
\ee

Due to the gauge fixing condition~(\ref{Sfix1}), $h_I\lmd^I=h_I\chi_S^I=0$. 
Thus, these combinations should be interpreted as the fermionic components 
of the graviphoton multiplet whose only nonvanishing component is 
the graviphoton~$B_\mu$. 
Note that $h^x_I\equiv g^{xy}a_{IJ}h^J_y$ are orthogonal to $h_I$. 
Thus, the physical fermions are specified as 
\bea
 \tl{\lmd}^x \defa h^x_I\lmd^I=\frac{M_5^{1/2}}{\sqrt{2}}\lmd^{I=x}
 +\cO(M_5^{-5/2}), \nonumber\\
 \tl{\chi}_S^x \defa h^x_I\chi_S^I=\frac{M_5^{1/2}}{\sqrt{2}}\chi_S^{I=x}
 +\cO(M_5^{-5/2}). 
 \label{phys_fmn}
\eea

Therefore, we can express the physical superfields in terms of 
the original gauge superfields as 
\bea
 \tl{V}^x \eql \frac{M_5^{1/2}}{\sqrt{2}}V^{I=x}+\cO(M_5^{-5/2}), 
 \nonumber\\
 \tl{\Phi}_S^x \eql \frac{M_5^{1/2}}{\sqrt{2}}\Phi_S^{I=x}+\cO(M_5^{-5/2}).  
\eea
Terms of $\cO(M_5^{-5/2})$ are cubic or higher order for the physical fields. 

The graviphoton~$B_\mu$ is identified with $W^{I=0}_\mu$ 
at the leading order of $M_5^{-1}$-expansion in our case. 
Since we are not interested in the fluctuation modes 
of the gravitational multiplet in this paper, 
we will replace the graviphoton by its VEV in the following.

\subsection{Deviation from BPS limit}
5D SUSY transformation is parametrized by an $\SUu$ Majorana spinor~$\ep^i$ 
($i=1,2$). 
This can be rewritten as two 4D Majorana spinors. 
\bea
 \ep^+ \defa \cP_{\rm R}\ep^1+\cP_{\rm L}\ep^2, \nonumber\\
 \ep^- \defa i\brkt{\cP_{\rm R}\ep^2+\cP_{\rm L}\ep^1}, 
\eea
where the projection operators~$\cP_{\rm R,L}$ are defined 
in Eq.(\ref{prj_op}). 
The signs in the superscripts denote the corresponding parity 
under the orbifold transformation. 
Thus, after the orbifold projection, 5D $\cN=1$ SUSY is broken to 
4D $\cN=1$ SUSY parametrized by $\ep^+$. 
In the following, we will focus on this SUSY and its breaking. 

Now, let us define the following quantities. 
\bea
 \fG \defa \gey^{-1}\dot{\sgm}-\frac{2}{3M_5^3}\vev{\cN_IY^{I(3)}}, \nonumber\\
 \fh^\alp \defa \gey^{-1}\Lvev{\der_y\cA_1^\alp+i\tl{\cF}_1^\alp
  -\brkt{W_y^I-i\gey M^I}(gt_I)^\alp_{\;\;\bt}\cA_1^\bt
  +\frac{3}{2}\dot{\sgm}\cA_1^\alp}, \nonumber\\
 \DV^I \defa -\gey^{-1}\Lvev{\der_yM^I-2\gey Y^{I(3)}
 +\dot{\sgm}M^I}, \nonumber\\
 \fS^I \defa -i\vev{Y^{I(1)}+iY^{I(2)}}. 
 \label{odpmt}
\eea
Here, we have assumed that the graviphoton~$W_y^0$ 
does not have a non-zero VEV. 
Note that $\fh^\alp$, $\DV^I$ and $\fS^I$ are proportional to 
VEVs of the auxiliary fields in $\Phi^\alp$, $V^I$ and $\Phi_S^I$ 
in the case of $\vev{V_y^{(1)}+iV_y^{(2)}}=0$.  
Using these quantities, 
VEVs of SUSY variations for fermions can be written as 
\bea
 \vev{\dQ\psi_{m{\rm R}}^1} \eql -\frac{e^\sgm}{3M_5^3}\cN_I\bar{f}_S^I 
 (\gm_{\udl{m}}\ep^+)_{\rm R}, \;\;\;\;\; 
 \vev{\dQ\psi_{m{\rm R}}^2} = -\frac{i}{2}e^\sgm\fG
 (\gm_{\udl{m}}\ep^+)_{\rm R}, 
 \nonumber\\
 \vev{\dQ\psi_{y{\rm R}}^1} \eql \frac{\gey}{2}\fG\ep^+_{\rm R}, \;\;\;\;\; 
 \vev{\dQ\psi_{y{\rm R}}^2} = -i\brkt{\vev{V_y^{(1)}+iV_y^{(2)}}
 +\frac{\gey}{3M_5^3}\vev{\cN_I}\fS^I}\ep^+_{\rm R}, \nonumber\\
 \vev{\dQ\ztR^\alp} \eql i\brkt{\fh^\alp-\frac{3}{2}\fG\vev{\cA_1^\alp}
  +\frac{i}{M_5^3}\fS^I\vev{\cN_I\cA_2^\alp}}\ep_{\rm R}^+, \nonumber\\
 \vev{\dQ\Omg_{\rm R}^{I1}} \eql \frac{i}{2}\brkt{\DV^I+\vev{M^I}\fG}
 \ep^+_{\rm R},  \;\;\;\;\; 
 \vev{\dQ\Omg_{\rm R}^{I2}} = \brkt{-\fS^I+\frac{1}{3M_5^3}\vev{M^I\cN_J}\fS^J}
 \ep^+_{\rm R}, 
 \label{SUSYvar}
\eea
where the transformation parameter~$\ep^+$ is assumed as 
\be
 \ep^+(y) = e^{\frac{\sgm(y)}{2}}\ep^+_0. \;\;\;\;\;
 (\ep^+_0\mbox{: 4D Majorana constant spinor})
\ee
Thus, the quantities defined in Eq.(\ref{odpmt}) 
and $\vev{V_y^{(1)}+iV_y^{(2)}}$ 
characterize the deviation from the BPS limit. 
However, not all of them are independent quantities. 
In fact, there are some relations among them. 
By taking the SUSY variations of the gauge fixing condition~(\ref{Sfix1}) 
and using Eq.(\ref{SUSYvar}), 
we will obtain a relation
\be
 \vev{\cN_I}\brkt{\DV^I+\vev{M^I}\fG}=0. \label{od_rel1}
\ee
Similarly, from the gauge fixing condition~(\ref{Sfix2}), 
we will obtain 
\bea
 \fh^1 \eql \vev{\cA^2_2}^{-1}\brc{\frac{3}{2}M_5^3\fG
 -\sum_{\udl{\alp},\udl{\bt}=3}^{2\nH+2}\rho_{\udl{\alp}\udl{\bt}}
 \vev{\cA_2^{\udl{\alp}}}\fh^{\udl{\bt}}}, \nonumber\\
 \fh^2 \eql -\vev{\cA_2^2}^{-1}\brc{\sum_{\udl{\alp}=3}^{2\nH+2}
 \vev{\bar{\cA}_2^{\udl{\alp}}}\fh^{\udl{\alp}}+i\vev{\cN_I}\fS^I}, 
 \label{od_rel2}
\eea
where 
\be
 \vev{\cA_2^2}=\brkt{M_5^3+\abs{\vev{\cA_2^3}}^2
 +\abs{\vev{\cA_2^4}}^2}^{1/2}. 
\ee

\section{Action after gauge fixing} \label{EFT} 
In this section, we will derive the expression 
after imposing the gauge fixing conditions~(\ref{Dfix1}-\ref{Sfix2}) 
on the invariant action~(\ref{Sinv}). 

\subsection{Brane action} \label{action_fml}
Before proceeding to the derivation of the bulk action, 
let us discuss the brane action briefly. 
In addition to the bulk action, we can introduce the brane actions 
localized at the fixed boundaries of the orbifold. 
If we neglect fluctuations of the gravitational fields, 
the 4D superconformal invariant action~\cite{KU} can be expressed as follows. 
\bea
 S_{\rm brane} \eql \sum_{\hat{y}=0,\pi R}
 \int\dr^5x\;c_{\hat{y}}\dlt(y-\hat{y})
 \cL_{\rm brane}^{(\hat{y})}, \nonumber\\
 \cL_{\rm brane}^{(\hat{y})} \eql \brc{\int\dr^2\tht\;
 f_{\bar{I}\bar{J}}(S)\cW^{\bar{I}}\cW^{\bar{J}}+\hc} 
 \nonumber\\
 &&-e^{2\sgm}\int\dr^4\tht\;\bar{\Sgm}_{\hat{y}}\Sgm_{\hat{y}} 
 e^{-K(S,\bar{S})}
 +e^{3\sgm}\brc{\int\dr^2\tht\;\Sgm_{\hat{y}}^3 P(S)+\hc}, 
\eea
where $\hat{y}=0,\pi R$ denote the coordinates of the fixed boundaries, 
and $c_{\hat{y}}$ are some dimensionless constants which are assumed 
as small numbers. 
$f_{\bar{I}\bar{J}}$, $K$ and $P$ are the gauge kinetic functions, 
the K\"{a}hler potential, and the superpotential, respectively. 
The indices~$\bar{I},\bar{J}$ run over not only the brane localized 
vector multiplets but also induced ones on the boundaries 
from the bulk multiplets. 
The chiral superfield~$\Sgm_{\hat{y}}=\vph_{\hat{y}}^0
-\tht\chi_{\hat{y}}^0-\tht^2\cF_{\hat{y}}^0$ 
is a 4D compensator superfield, and $S^a$ are chiral matter superfields. 
The Weyl weights of the lowest components in $\Sgm_{\hat{y}}$ and $S^a$ 
are one and zero, respectively. 
The warp factors in $\cL_{\rm brane}^{(\hat{y})}$ come from 
the induced metric on the boundaries. 
Note that the 4D compensator~$\Sgm_{\hat{y}}$ must be induced 
from the 5D compensator because the gravity is unique. 
Since the 5D compensator scalar~$\cA_2^{\alp=2}$ has the Weyl weight $3/2$, 
we should identify $\Sgm_{\hat{y}}$ as\footnote{
We cannot use $\Phi^{\alp=1}$ because it is odd under the orbifold parity 
and vanishes on the boundaries. } 
\be
 \Sgm_{\hat{y}} = \Sgm|_{y=\hat{y}}, 
\ee
where
\be
 \Sgm = (\Phi^{\alp=2})^{2/3}. 
\ee
Chiral matter fields $S^a$ appearing in $S_{\rm brane}$ 
can contain the induced multiplets on the boundary 
from the 5D bulk multiplets. 
However, since the Weyl weight of $S^a$ must be zero, 
the physical bulk fields~$\Phi^\alp$ ($\alp\geq 3$) can appear 
in $S_{\rm brane}$ only in the form of 
\be
 S_{\hat{y}}^v \equiv M_5^{-1/2}H^v|_{y=\hat{y}}=
 \vph_{\hat{y}}^v-\tht\chi_{\hat{y}}^v-\tht^2\cF_{\hat{y}}^v, 
\ee
where $v=1,2,\cdots,\nH$, and 5D superfields~$H^v$ are defined by 
\be
 H^v \equiv \frac{\sqrt{2}M_5^{3/2}\Phi^{\alp=2v+2}}{\Phi^{\alp=2}} 
 =h^v-\tht\chi_H^v-\tht^2\cF_H^v.
\ee
The numerical factor is for the canonical normalization in the bulk action. 
Similarly, we will define another bulk fields~$H^{Cv}$ ($v=1,2,\cdots,\nH$) as 
\be
 H^{Cv} \equiv \frac{\sqrt{2}M_5^{3/2}\Phi^{\alp=2v+1}}{\Phi^{\alp=2}} 
 =h^{Cv}-\tht\chi_H^{Cv}-\tht^2\cF_H^{Cv}. 
\ee
Note that $H^{Cv}$ are odd under the orbifold parity and vanish 
on the boundaries, while $H^v$ are even.  
(See Table~\ref{Z2_parity}.)

The gauge fixing conditions~(\ref{Dfix2}) and (\ref{Sfix2}) are rewritten 
on the boundary in terms of the components of $\Sgm_{\hat{y}}$ 
and $S_{\hat{y}}^v$ as 
\bea
 \vph_{\hat{y}}^0 \eql M_5\brkt{1+\frac{1}{6M_5^2}
 \sum_{v=1}^{\nH}(\abs{\vph_{\hat{y}}^v}^2+\abs{\vph_{\hat{y}}^{Cv}}^2)
 +\cO(M_5^{-4})}, 
 \nonumber\\
 \chi_{\hat{y}}^0 \eql \frac{1}{3M_5^2}\vph_{\hat{y}}^0
 \sum_{v=1}^{\nH}(\bar{\vph}_{\hat{y}}^v\chi_{\hat{y}}^v
 +\bar{\vph}_{\hat{y}}^{Cv}\chi_{\hat{y}}^{Cv})+\cO(M_5^{-4}). 
\eea
where $\cO(M_5^{-4})$ terms are quartic and higher order terms 
for the physical fields. 
These conditions coincide with the gauge fixing conditions 
in the superconformal formulation of 4D supergravity 
up to the quadratic order for the physical fields, 
except for the overall factor of $\vph_{\hat{y}}^0$.
(See the appendix of Ref.~\cite{geoSB4D}.)
The discrepancy of the overall factor leads to a wrong coefficient 
of the Einstein-Hilbert term. 
However, this is not a serious problem. 
Strictly speaking, when we add the brane actions to the bulk action, 
the $\bdm{D}$-gauge fixing condition~(\ref{Dfix1}) and (\ref{Dfix2}) 
should be modified so that the 4D Einstein-Hilbert term 
is canonically normalized after integrating out $y$-coordinate. 
However, this modification is almost negligible because of 
the suppression by the large volume of the extra dimension 
and the smallness of $c_{\hat{y}}$ in our case. 
The coefficient of the 4D Einstein-Hilbert term is determined 
mainly by the bulk term. 
Thus, the gauge fixing conditions~(\ref{Dfix1})-(\ref{Sfix2}) 
can also be applied for the bulk-brane system.

\subsection{Bulk action}
Now we will derive the bulk action~$S_{\rm bulk}$ after the gauge fixing. 
For simplicity, we will assume that the compensator multiplet 
and the hidden sector fields~$\Phi^\alp$ ($\alp=1,\cdots,4$) are charged 
under only the graviphoton~$W^0_\mu$. 
All the directions of the gauging are chosen to $\sgm_3$-direction 
since the gauging along the other directions mixes 
$\Phi^{2\hat{\alp}+1}$ and $\Phi^{2\hat{\alp}+2}$, 
which have opposite parity eigenvalues. 
Namely, 
\be
 gt_0=-i\brkt{\begin{array}{ccc} g_{\rm c}^0 & & \\ 
 & g_{\rm h}^0 & \\ & & \bdm{g}^0 
 \end{array}}\otimes \sgm_3, 
 \;\;\;
 gt_I=-i\brkt{\begin{array}{ccc} 0 & & \\ & 0 & \\ 
 & & \bdm{g}^I \end{array}}\otimes \sgm_3,  \;\;\;(I\neq 0)
 \label{gauge_cp}
\ee
where $\bdm{g}^I\equiv\diag(g^I_2,g^I_3,\cdots,g^I_{\nH})$ 
($I=0,1,\cdots,\nV$) are $(\nH-1)\times(\nH-1)$ matrices 
of the gauge couplings for the hypermultiplets in the visible sector. 
Note that the gauge couplings~$(g_{\rm c}^0, g_{\rm h}^0, \bdm{g}^0)$ are odd 
under the orbifold parity 
since $\Pi_{I=0}=-1$ in Table~\ref{Z2_parity}. 
Namely, $(g_{\rm c}^0, g_{\rm h}^0, \bdm{g}^0)$ have kink profiles.\footnote{
The kink-type gauge couplings can be realized in SUGRA context 
by the mechanism proposed in Ref.~\cite{BKV}. }
The other gauge couplings are even under the parity 
and constant in the whole range of $y$. 

Since we are interested only in the visible sector, 
we will neglect the fluctuation modes of the stabilizer 
hypermultiplet~$(\Phi^{\alp=3},\Phi^{\alp=4})$ around the background. 
Then, the bulk action is written as 
\bea
 S_{\rm bulk} \eql \int\dr^5x\;\cL_{\rm SF}+\cL_{\rm SB}, \nonumber\\
 \cL_{\rm SF} \eql \brc{\int\dr^2\tht\;\frac{1}{4}T\tl{\cW}^x\tl{\cW}^x+\hc}
 +\frac{e^{2\sgm}}{2}\int\dr^4\tht\;\frac{T+\bar{T}}{V_T^2}
 \brkt{\der_y\tl{V}^x+i\tl{\Phi}_S^x-i\bar{\tl{\Phi}}_S^x}^2 \nonumber\\
 &&-e^{2\sgm}\int\dr^4\tht\;V_T(\bar{\Sgm}\Sgm)^{3/2}
 \brc{2-\frac{1}{M_5^3}\brkt{\bar{H}e^{2\tlg^x\tl{V}^x}H
 +\bar{H}^Ce^{-2\tlg^x\tl{V}^x}H^C}}
 \nonumber\\
 &&+e^{3\sgm}\brc{\frac{1}{M_5^3}\int\dr^2\tht\;\Sgm^3 H^C
 \brkt{\frac{1}{2}\lrder_y+m_0T-2i\bdm{\tl{g}}^x\tl{\Phi}_S^x}H+\hc}, 
 \nonumber\\
 \cL_{\rm SB} \eql -e^{4\sgm}\gey\brc{\hat{O}_h(|h|^2+|h^{C}|^2)
 +\fG\brkt{1+\frac{\abs{\cA_2^3}^2+\abs{\cA_2^4}^2}{M_5^3}}
 (\bar{h}m_0h-\bar{h}^Cm_0h^C)} \nonumber\\
 &&-e^{4\sgm}\gey\hat{O}_\phi\brkt{(\phi^x)^2} 
 -e^{2\sgm}\fG\brc{\gey\phi^x\tl{D}^x
 -\frac{1}{2}(\tl{\chi}_S^x\tl{\lmd}^x+\hc)}+\cdots,  
 \label{gn_fml}
\eea
where $H^v$, $H^{Cv}$, and $\Sgm$ are defined in the previous subsection. 
The index~$v=2,3,\cdots,\nH$ for the hypermultiplets is suppressed, and 
summations for the indices~$x=1,2,\cdots,\nV$ are implicit. 
The derivative operator~$\lrder_y$ is defined as 
$A\!\!\lrder_y\!\!B\equiv A\der_y B-B\der_y A$. 
The mass matrix~$m_0$ and the physical (dimensionful) gauge 
couplings~$\bdm{\tl{g}}^x$ are defined as 
\be
 m_0 \equiv M_5 \bdm{g}^0, \;\;\;\;\;
 \bdm{\tl{g}}^x \equiv \frac{\sqrt{2}}{M_5^{1/2}}\bdm{g}^{I=x}, 
\ee
and the superfield~$T$ and the operators~$\hat{O}_h$ and $\hat{O}_\phi$ 
are 
\bea
 T \defa -\frac{2i}{M_5}\Phi_S^0, \nonumber\\
 \hat{O}_h \defa \brkt{1+\frac{\abs{\cA_2^3}^2+\abs{\cA_2^4}^2}{M_5^3}}
 \brc{g_{\rm c}^0\DV^0-\frac{\fh^1}{M_5^{3/2}}\gey^{-1}
 \brkt{\der_y+\gey M_5 g_{\rm c}^0+\frac{3}{2}\dot{\sgm}}} \nonumber\\
 &&-\frac{2\fh^1}{M_5^{9/2}}\gey^{-1}
 \Re(\bar{\cA}_2^3\der_y\cA_2^3+\bar{\cA}_2^4\der_y\cA_2^4), 
 \nonumber\\
 \hat{O}_\phi \defa \brc{-\frac{\fG}{2}\gey^{-1}\brkt{\der_y+2\dot{\sgm}}
 +\frac{\fG^2}{2}+\frac{2}{M_5^2}\abs{\fS^0}^2-\frac{4g_{\rm h}^0}{3M_5^2}
 \brkt{\cA_2^3\fh^4-\cA_2^4\fh^3}
 -\frac{4g_{\rm c}^0}{3M_5^{1/2}}\fh^1}. \nonumber\\
\eea
Here, we have used the relation 
\be
 \DV^0=-M_5 \fG, \label{od_rel1p}
\ee
which is derived from Eq.(\ref{od_rel1}) under the assumption 
that $\vev{\phi^x}=0$.\footnote{
In this paper, we assume that only $\cA_2^{\alp=3,4}$ have 
nontrivial vacuum configurations. } 
The ellipsis in Eq.(\ref{gn_fml}) denotes 
quartic or higher order terms for the physical fields. 
Such terms are suppressed by the large 5D Planck mass~$M_5$ and 
are negligible. 
Field independent terms are dropped here. 

In the above expression, fermionic components and fluctuation modes of 
auxiliary fields in $\Sgm$ and $T$ 
only contribute $\cO(\kp)$ quartic or higher order terms 
that are neglected here. 
Thus, in Eq.(\ref{gn_fml}), $\Sgm$ and $T$ can be understood as 
the following spurion superfields. 
\bea
 \Sgm \eql \brkt{M_5+\frac{\abs{\vev{\cA_2^3}}^2
 +\abs{\vev{\cA_2^4}}^2}{3M_5^2}}
 -\tht^2\cF_\Sgm, \nonumber\\
 T \eql \brkt{\gey-\frac{2i}{M_5}\vev{W_y^0}}-\tht^2\cF_T, \label{spurions}
\eea
where $\cF_T\equiv -2ie^\sgm\brkt{\vev{V_y^{(1)}+iV_y^{(2)}}
+\gey M_5^{-1}\fS^0}$. 
From Eqs.(\ref{def_VT}) and (\ref{spurions}), 
we can see that $T$ and $V_T$ correspond to 
the radion superfield. 
We will discuss it in the next section. 
All the SUSY breaking terms in $\cL_{\rm SB}$ involve quantities 
defined in Eq.(\ref{odpmt}) and thus vanish in the case of 
a BPS background. 

\subsection{Radion superfield}
Here, we will consider the case that the background is BPS. 
In this case, $\cL_{\rm SB}$ in Eq.(\ref{gn_fml}) vanishes and 
the action can be expressed by only superfields. 
Besides, $V_T$ can be expressed by $T$ as\footnote{
This relation is also mentioned in Ref.~\cite{CST}.} 
\be
 V_T = \frac{T+\bar{T}}{2}, 
 \label{VT-T_rel}
\ee
where $\cF_T=-2ie^\sgm\vev{V_y^{(1)}+iV_y^{(2)}}$ in $T$. 
The above relation always holds only if $\fS^0=0$, \ie, 
only one of $\cA_2^{\alp=3}$ and $\cA_2^{\alp=4}$ has a nonzero vacuum 
configuration. (See Eqs.(\ref{odpmt}) and (\ref{Y12}).) 
Here, we will rescale the compensator superfield as 
\be
 \Sgm^\prime \equiv \frac{\Sgm}{M_5}=
 \brkt{1+\frac{\abs{\vev{\cA_2^3}}^2+\abs{\vev{\cA_2^4}}^2}{3M_5^3}}
 -\tht^2\cF^\prime_\Sgm. 
\ee
Then, the superfield Lagrangian~$\cL_{\rm SF}$ becomes 
\bea
 \cL_{\rm SF} \eql \brc{\int\dr^2\tht\;\frac{1}{4}T\tl{\cW}^x\tl{\cW}^x+\hc}
 +e^{2\sgm}\int\dr^4\tht\;\frac{2}{T+\bar{T}}
 \brkt{\der_y\tl{V}^x+i\tl{\Phi}_S^x-i\bar{\tl{\Phi}}_S^x}^2 \nonumber\\
 &&-e^{2\sgm}\int\dr^4\tht\;\frac{T+\bar{T}}{2}
 (\bar{\Sgm}^\prime\Sgm^\prime)^{3/2}
 \brc{2M_5^3-\bar{H}e^{2\tlg^x\tl{V}^x}H-\bar{H}^Ce^{-2\tlg^x\tl{V}^x}H^C}
 \nonumber\\
 &&+e^{3\sgm}\brc{\int\dr^2\tht\;\Sgm^{\prime 3} H^C
 \brkt{\frac{1}{2}\lrder_y+m_0T-2i\bdm{\tl{g}}^x\tl{\Phi}_S^x}H+\hc}. 
\eea
In the limit that $\Lmd_{\rm hid}/M_5 \to 0$, \ie, 
$\vev{\cA_2^{\alp=3,4}}/M_5^{3/2}\to 0$, 
the warp factor is calculated as  
\be
 \sgm(y)=-\frac{2\gey M_5 g_{\rm c}^0}{3}y, \label{wp_fct}
\ee
by solving the Killing spinor equation. 
Here, the normalization of the warp factor is chosen as $\sgm(0)=0$. 
This is the case of the supersymmetric Randall-Sundrum model.\footnote{
Note that the gauge coupling~$g_{\rm c}^0$ is odd under the orbifold parity. }
In this case, the above result becomes very similar 
to the result of Ref.~\cite{marti}, 
except for the dependence of the {\it radion superfield}~$T$ 
in the warp factor.  
We can reproduce the $T$-dependence in the warp factor of Ref.~\cite{marti} 
by the following redefinition. 
\bea
 V_T &\to& \exp\brc{-\brkt{\frac{T+\bar{T}}{2\gey}-1}\sgm}V_T, 
 \nonumber\\
 \Sgm^\prime &\to& \exp\brc{\brkt{\frac{T}{\gey}-1}\sgm}\Sgm^\prime. 
\eea
Note that this redefinition does not change the lowest components of 
$V_T$ and $\Sgm^\prime$, but change only their auxiliary fields.\footnote{
We have assumed that $\vev{W_y^0}=0$. }
In this case, however, the relation~(\ref{VT-T_rel}) no longer holds, 
and we cannot reproduce the $T$-dependence of Ref.~\cite{marti} 
in the parts of $d^4\tht$-integration in the action. 

The reason for the discrepancy in the $T$-dependence between our action 
and that of Ref.~\cite{marti} is as follows. 
As we can see from Eq.(\ref{wp_fct}), 
the warp factor~$e^\sgm$ has an $\gey$-dependence. 
In Ref.~\cite{marti}, all $\gey$ appearing in the action are promoted 
to the radion superfield~$T$. 
However, note that the $\gey$-dependence of the warp factor 
is induced by solving the equation of motion, 
or the Killing spinor equation in the BPS case. 
In other words, $\sgm(y)$ is independent of $\gey$ 
at the original SUGRA action. 
($\ge{m}{\udl{n}}$ is independent of $\ge{y}{4}$.)
All $\gey$ except for those in the warp factor should be incorporated 
into the superfields because they originate from the f\"{u}nfbein 
in the superconformal invariant action. 
On the other hand, $\gey$ in the warp factor appears only after 
substituting the background solution into the action. 
Therefore, this $\gey$ should not be accompanied 
by the auxiliary field~$\cF_T$.

\subsection{Comment on Scherk-Schwarz breaking} \label{SSSB}
Next, we will briefly comment on the relation 
between nonzero $\cF_T$ and Scherk-Schwarz (SS) SUSY breaking~\cite{SS} 
for further understanding of the radion superfield. 
In the framework of superconformal gravity, 
twisting boundary conditions for fields with $SU(2)_R$ indices 
can be realized by twisting only the compensator gauge fixing as 
\be
 \cA_i^a = \brkt{e^{i\alp(y)\vec{\omg}\cdot\vec{\sgm}}}_{\;\;i}^a
 \brkt{M_5^3+\sum_{\udl{\alp}=3}^{2\nH}\abs{\cA_2^{\udl{\alp}}}^2}^{1/2}, 
 \;\;\;\;\; (a=1,2)
\ee
where $\vec{\omg}=(\omg^1,\omg^2,\omg^3)$ is a unit twist vector, 
while the physical fields are kept untwisted. 
In fact, we can move to the usual SS twisting for the physical fields 
by rotating back the above gauge fixing to the usual one~(\ref{Ufix})
with an $\SUu$ matrix. 
For the single-valuedness of the compensator fields, 
the twisting parameter~$\alp(y)$ satisfies 
\be
 \alp(y+2\pi R)=\alp(y)+2n\pi. \;\;\;\;\;(\mbox{$n$: nonzero integer}) 
 \label{prd_alp}
\ee
Notice that we can also move to the untwisted compensator 
by $SU(2)_C$ rotation mentioned in Sect.\ref{GFcond}, 
\be
 \cA_i^a \to U^a_{\;\;b}\cA^b_i, \;\;\;\;\;
 \zeta^a \to U^a_{\;\;b}\zeta^b, 
\ee
where $U^a_{\;\;b}$ ($a,b=1,2$) is an $SU(2)_C$ matrix defined as 
\be
 U^a_{\;\;b} \equiv \brkt{e^{-i\alp(y)\vec{\omg}\cdot\vec{\sgm}}}^a_{\;\;b}. 
\ee
From Eq.(\ref{Y12}), we can see that 
the $\SUu$ gauge fields~$V_y^{(r)}$ ($r=1,2,3$) 
are shifted in this basis as
\be
 V_y^{(r)} \to V_y^{(r)}+\frac{\der_y\alp\abs{\cA_2^2}^2}{M_5^3}\omg^r 
 =V_y^{(r)}+\der_y\alp\brkt{1+\frac{1}{M_5^3}\sum_{\udl{\alp}=3}^{2\nH}
 \abs{\cA_2^{\udl{\alp}}}^2}\omg^r, 
\ee
where we have used the gauge fixing conditions in the second equality. 
Thus, the SS twisting induces the non-vanishing Wilson line of 
the $\SUu$ gauge field. 
\be
 \int_0^{2\pi R}dy\;\vev{V_y^{(r)}}=2\pi n(1+\dlt)\omg^r, 
\ee
where 
\be
 \dlt \equiv \int_0^{2\pi R}dy\;
 \frac{\der_y\alp(y)}{2\pi n M_5^3}
 \brkt{\abs{\vev{\cA_2^3}(y)}^2+\abs{\vev{\cA_2^4}(y)}^2}. 
\ee
We have used Eq.(\ref{prd_alp}) and assumed that the SS twisting 
is the only source of the non-vanishing Wilson line. 
In fact, the stabilizer fields~$\cA_2^{\alp=3,4}$ can also 
contribute to the Wilson line, but it is estimated as 
$\cO(\Lmd_{\rm hid}^3/M_5^3)$ from Eq.(\ref{Y12}) and negligible. 

Thus, the SS twisting is equivalent to the non-vanishing Wilson line. 
Here, notice that the consistency of the SS twisting 
with the orbifold transformation characterized by Table~\ref{Z2_parity} 
requires $\omg^3=0$.\footnote{
In general, the orbifold transformation can mix two components 
of an $\SUu$-doublet spinor~$\psi^i$ as 
$\psi^i(-y)=\Pi\gm_5 M^i_{\;\;j}\psi^j(y)$, 
where the mixing matrix~$M^i_{\;\;j}$ satisfies $M^*=-\sgm_2 M\sgm_2$. 
We have chosen as $M=\sgm_3$ in this paper. }
Thus, we can ignore the dependence of $V_y^{(3)}$ in the action, as we did. 
As mentioned in the previous subsection, the auxiliary field of radion 
is provided by $\cF_T=-2ie^\sgm\vev{V_y^{(1)}+iV_y^{(2)}}$. 
Therefore, we can conclude that the SS SUSY breaking is equivalent 
to the breaking due to a nonzero~$\cF_T$, as is well-known 
in the case of flat spacetime~\cite{marti}. 

According to Ref.~\cite{HNOO}, however, the twisted boundary condition 
leads to inconsistent theory in the slice of AdS spacetime. 
In the superconformal formulation, 
this can be seen from the appearance of a nonzero mass 
of the graviphoton~$B_\mu$, which is proportional to $\vec{\omg}$ and 
$g_{\rm c}^0$. 
Thus, this inconsistency of the theory can be avoided 
if $g_{\rm c}^0=0$ and the warp factor can be generated only from 
the vacuum configuration of the bulk scalars. 


\subsection{Geometry mediated SUSY breaking} \label{geoSB}
Finally, we will discuss the SUSY breaking effect induced by 
the non-BPS background geometry. 
In order to focus on this effect, we will consider the case that 
there is no SS twisting discussed in the previous subsection. 
In this case, SUSY breaking is characterized by the quantities defined in 
Eq.(\ref{odpmt}). 
Among them, only $\fh^{3,4}$ and $\fG$ are independent in our case. 
In fact, $\fh^{1,2}$ and $\DV^0$ are related to them 
through Eqs.(\ref{od_rel2}) and (\ref{od_rel1p}), and the others are zero. 
Here, we assume that only one complex scalar~$\cA_2^3$ or $\cA_2^4$ 
has a nontrivial field configuration so that $\fS^0=0$.\footnote{
One example of such non-BPS configuration is found in Ref.~\cite{EMS2} 
in the 5D global SUSY model. 
The authors insist that their solution can be embedded into 5D SUGRA. } 
The quantities~$\fh^{3,4}$ and $\fG$ characterize SUSY breaking 
in the stabilization sector and 
the deviation of the geometry from the BPS limit, respectively. 
They are calculated immediately once the non-BPS background is found. 

The action is calculated as 
\bea
 S \eql \int\dr^5x\;\cL_{\rm SF}+\cL_{SB}, \nonumber\\
 \cL_{\rm SF} \eql \brc{\int\dr^2\tht\;\frac{1}{4}\tl{\cW}^x\tl{\cW}^x+\hc}
 +e^{2\sgm}\int\dr^4\tht\;\brkt{\der_y\tl{V}^x+i\tl{\Phi}_S^x
 -i\bar{\tl{\Phi}}_S^x}^2 \nonumber\\
 &&+e^{2\sgm}\int\dr^4\tht\;\brkt{\bar{H}e^{2\tlg^x\tl{V}^x}H
 +\bar{H}^C e^{-2\tlg^x\tl{V}^x}H^C} \nonumber\\
 &&+e^{3\sgm}\brc{\int\dr^2\tht\; H^C\brkt{
 \frac{1}{2}\lrder_y+m_0-2i\bdm{\tl{g}}^x\tl{\Phi}_S^x}H+\hc}, \nonumber\\
 \cL_{\rm SB} \eql e^{4\sgm}\fG\left\{(\der_y+3\dot{\sgm}
 +2M_5 g_{\rm c}^0-\fG)(\phi^x)^2 \right. \nonumber\\
 &&\hspace{13mm}+\brkt{\frac{3}{2}\der_y+\frac{5}{2}M_5 g_{\rm c}^0
 +\frac{9}{4}\dot{\sgm}}(|h|^2+|h^C|^2)
 -(\bar{h}m_0 h-\bar{h}^Cm_0 h^C) \nonumber\\
 &&\hspace{13mm}\left.+\frac{e^{-2\sgm}}{2}
 (\tl{\chi}_S^x\tl{\lmd}^x+\hc)
 +\phi^x(\bar{h}\bdm{\tl{g}}^x h-\bar{h}^C\bdm{\tl{g}}^x h^C)\right\}
 +\cdots, 
 \label{geoSBterms}
\eea
where the ellipsis denotes the negligible quartic or higher order terms
for the physical fields. 
Here, we have rescaled $y$ so that $\gey=1$, and used 
\be
 \fh^1 \simeq \frac{3}{2}M_5^{3/2}\fG, 
\ee
which follows from Eq.(\ref{od_rel2}) 
under the assumption~$\Lmd_{\rm hid}\ll M_5$. 
We have redefine the auxiliary field~$\tl{D}^x$ in $\tl{V}^x$ as 
$\tl{D}^x\to\tl{D}^x-2e^{2\sgm}\fG\phi^x$ 
in order to absorb the explicit dependence of $\tl{D}^x$ in $\cL_{\rm SB}$ 
of Eq.(\ref{gn_fml}). 

Note that all SUSY breaking terms are proportional to $\fG$, 
which characterizes the deviation of the warp factor~$\sgm(y)$ 
from the BPS limit. 
This means that SUSY breaking in the visible sector mainly comes from 
only the deformation of the background geometry. 
We can also see that no $A$-terms or $B$-terms for the matter hyperscalars 
are induced. 
Since the physical gauge scalar superfields~$\tl{\Phi}_S^x$ 
($x=1,2,\cdots,\nV$) are odd under the orbifold transformation, 
they have no zero-modes and decouple 
below the compactification scale~$R^{-1}$.\footnote{
Since $\Lmd_{\rm hid}^{-1}$ is the characteristic length of 
the nontrivial field configuration, the radius~$R$ is generally larger 
than $\Lmd_{\rm hid}^{-1}$. } 
Thus, the only SUSY breaking terms induced by the deformed geometry 
are the mass terms for the hyperscalars in the low-energy effective theory. 

We can check that the non-BPS bulk geometry does not cause 
SUSY breaking in the brane actions at tree level.  
This can easily be understood from the fact that $\fG(y)$ is an odd function 
under the orbifold parity and vanishes on the boundaries.

\section{Summary and comments} \label{summary}
We have proposed a practical method of deriving 5D SUGRA action 
on a general warped geometry using 4D superfields. 
Our method is based on the superconformal formulation of 5D SUGRA 
discussed in Refs.~\cite{KO3,KO1,KO2,Ohashi}. 
We have expressed an invariant action in 4D $\cN=1$ superspace 
in the case that fluctuation modes of the gravitational fields 
are neglected. 
This greatly simplifies calculations and makes the procedure transparent 
thanks to the well-known 4D superfield formalism. 
A significant advantage of our method is that we can deal with a 
{\it general warped geometry} including a non-BPS background 
in the superfield formalism. 

In the case of a non-BPS background, SUSY breaking terms appear 
after the superconformal gauge fixing. 
This might seem an explicit breaking of SUSY by the gauge fixing conditions. 
However, this is not the case. 
Notice that the genuine SUSY transformation~$\tl{\dlt}_Q$ is different from 
a fermionic transformation $\tl{\dlt}_Q^\prime$ 
generated by acting a differential 
operator~$\ep^+_{\rm R}\hat{Q}+\bar{\ep}^+_{\rm R}\bar{\hat{Q}}$ 
on each superfield, where 
\be
 \hat{Q}_\alp=\frac{\der}{\der\tht^\alp}
 -i\brkt{\sgm^{\udl{m}}\btht}_\alp\der_m, \;\;\;\;\;
 \bar{\hat{Q}}^{\dot{\alp}}=\frac{\der}{\der\btht_{\dot{\alp}}}
 -i\brkt{\bar{\sgm}^{\udl{m}}\tht}^{\dot{\alp}}\der_m.  
\ee
The genuine SUSY~$\tl{\dlt}_Q$ is defined as a combination of 
$Q$-transformation~$\dlt_Q$, the conformal supersymmetry~$\dlt_S$, 
and the special conformal boost~$\dlt_K$, 
\be
 \tl{\dlt}_Q(\ep) \equiv \dlt_Q(\ep)+\dlt_S(\eta(\ep))+\dlt_K(\xi_K(\ep)), 
\ee
where transformation parameters~$\eta^i$ and $\xi_K^{\udl{\mu}}$ 
are determined so that 
$\tl{\dlt}_Q$ preserves the superconformal gauge fixing conditions. 
(See Eq.(D.6) in the second paper of Ref.~\cite{KO3} 
for the explicit expressions of $\eta^i$ and $\xi_K^{\udl{\mu}}$.)
On the other hand, we have defined $\tl{\dlt}_Q^\prime$ so that 
it preserves the gravitational background~(\ref{grv_bgd}), \ie,\footnote{
In order to construct 4D superfields, 
the transformation parameter~$\ep^i$ is restricted to 
$\ep^+=e^{\frac{\sgm}{2}}\ep^+_0$ ($\ep^+_0$: 4D Majorana constant spinor).} 
\be
 \eta^i=\frac{i}{2}\dot{\sgm}\gm_5\ep^i, \;\;\;\;\; 
 \xi_K^{\udl{\mu}}=0.  \label{eta_xi}
\ee
Namely, the gauge fixing conditions break $\tl{\dlt}_Q^\prime$-symmetry, 
not the genuine SUSY~$\tl{\dlt}_Q$. 
The latter is broken by the non-BPS background, 
and thus this breaking is spontaneous. 
Notice that the difference between $\tl{\dlt}_Q$ and $\tl{\dlt}_Q^\prime$ is 
a choice of $\eta$ and $\xi_K$, and  
the superconformal transformations of $\cA_i^\alp$, $W_\mu^I$ and $M^I$ 
include neither $\eta$ nor $\xi_K$. 
Therefore, both transformations~$\tl{\dlt}_Q$ and $\tl{\dlt}_Q^\prime$ 
are identical for these fields. 
Since the superpartners of these fields are defined by 
$\tl{\dlt}_Q\cA_i^\alp$, $\tl{\dlt}_Q W_\mu^I$ and $\tl{\dlt}_Q M^I$, 
the superfields constructed by $\tl{\dlt}_Q^\prime$ describe 
the correct supermultiplets for {\it the on-shell fields}. 
All the transformations including $\eta$ or $\xi_K$ involve 
the auxiliary fields. 
This means that the deviation of $\tl{\dlt}_Q^\prime$ from $\tl{\dlt}_Q$ 
can be pushed into the definition of the auxiliary fields 
of the superfields. 
In fact, $\dot{\sgm}$ in Eqs.(\ref{sf_comp1}) and (\ref{sf_comp2}) 
corresponds to the contributions of $\eta$ in Eq.(\ref{eta_xi}). 


Our result can be compared with that of Ref.~\cite{marti} 
in the case of a BPS background. 
We saw that the radion superfield originates 
from two different kinds of superconformal multiplets in Ref.~\cite{KO2}. 
One is a 4D real general multiplet whose lowest component is $\ge{y}{4}$ 
and the other is a 4D chiral multiplet whose lowest component is 
$(W_y^0+i\ge{y}{4}M^0)/2$. 
On the gravitational background~(\ref{grv_bgd}), 
the corresponding superfield of the former multiplet 
is a spurion superfield~$V_T$ while the latter becomes 
the gauge scalar superfield~$\Phi_S^0$ for the graviphoton 
multiplet.\footnote{
As mentioned in Ref.~\cite{KO2}, both multiplets have nontrivial 
transformation properties under the full 5D superconformal transformation, and 
it is a hard task to find how they appear in 4D action formulae 
in a 5D superconformal invariant way. 
On the other hand, our superfield formalism is much easier 
to deal with because we focus on only the $\tl{\dlt}^\prime_Q$-symmetry, 
which is just part of the full superconformal symmetry. } 
Although these superfields have different origins, 
they describe almost the same degrees of freedom 
on the gravitational background,\footnote{
The fermionic component of $\Phi_S^0$ only contributes to 
quartic or higher order terms suppressed by the 5D Planck mass~$M_5$, 
which are neglected in this paper. 
} except for $Y^{0(1)}+iY^{0(2)}$, 
which has a vanishing VEV if only one scalar of the stabilizer hypermultiplet 
has a nontrivial configuration. 
In such a case, both $V_T$ and $\Phi_S^0$ are expressed by a single chiral 
spurion field~$T$, which is commonly called the {\it radion superfield}. 
Since we started from the off-shell SUGRA formulation, 
we can derive the dependence of $T$ in the action without any ambiguity. 
Our result is slightly different from that of Ref.~\cite{marti} 
by the $T$-dependence in the warp factor. 
This discrepancy stems from the fact that 
the $\gey$-dependence of the warp factor is induced by solving 
the equation of motion (or the Killing spinor equation in the BPS case). 
In other words, $\sgm(y)$ is independent of $\gey$ 
at the original SUGRA action because $\ge{m}{\udl{n}}$ is independent of 
$\ge{y}{4}$. 
Therefore, $\gey$ in the warp factor should not be accompanied 
by the auxiliary field~$\cF_T$. 
Using our identification of the radion superfield, we can explicitly see that 
SUSY breaking by the auxiliary field~$\cF_T$ of the radion superfield 
is equivalent to the Scherk-Schwarz breaking, 
as is well-known in flat spacetime.  

We also discussed SUSY breaking induced by a deformed geometry. 
Here, we have assumed that the deformed (non-BPS) geometry comes from 
the backreaction of a nontrivial scalar configuration 
which is relevant to the radius stabilization. 
In the case that the stabilization sector is decoupled 
from our visible sector, all SUSY breaking effects are induced through 
the background geometry. 
In fact, 
the dominant SUSY breaking terms are proportional to $\fG$, 
which characterizes the deformation of the geometry. 
(See Eq.(\ref{geoSBterms}).) 
Namely, $\fG$ is a $y$-dependent order parameter of SUSY breaking 
in this case. 
We can see that the deformed geometry only induces the masses 
for the bulk scalars besides the scalar trilinear coupling involving 
the gauge scalars~$\phi^x$. 
It should also be noted that the geometry mediated SUSY breaking 
induces no SUSY breaking effects on the boundaries at tree level 
because the order parameter~$\fG(y)$ is an odd function 
under the orbifold parity and vanishes on the boundaries. 

Finally, we would like to comment on the recently appeared preprint~\cite{CST}
related to our work.  
In Ref.~\cite{CST}, the authors express 5D superconformal gravity action 
in terms of the action formulae for 4D superconformal gravity. 
Our action before the superconformal gauge fixing should be obtained 
by fixing the gravitational background in their action. 
The radion spurion superfield discussed in our paper is consistent 
with their identification of the radion multiplet. 
In our paper, we further show the action after the gauge fixings explicitly 
which is useful to derive the effective theory 
even on a non-BPS background (general warped background).

\vspace{3mm}

\begin{center}
{\bf Acknowledgments}
\end{center}
The authors would like to thank Ian-Woo~Kim and Keisuke~Ohashi 
for useful discussions. 
H.~A. is supported by KRF PBRG 2002-070-C00022. 
Y.~S. is supported from the Astrophysical Research Center 
for the Structure and Evolution of the Cosmos (ARCSEC) 
funded by the Korea Science and Engineering Foundation 
and the Korean Ministry of Science. 

%

\appendix

\section{Notation} \label{notation}
Basically, we follow the notation of Refs.~\cite{KO1,KO2}. 
The metric convention is 
\be
 \eta_{\mu\nu}=\diag(1,-1,-1,-1,-1). 
\ee

We choose the following representation for $\gm$-matrices. 
\be
 \gm^0 = \mtrx{}{-1}{-1}{}, \;\;\;\;\;
 \gm^k = \mtrx{}{\sgm_k}{-\sgm_k}{}. \;\;\;\;\;
 \gm^4 =\mtrx{-i}{}{}{i},  
\ee
where $\sgm_k$ ($k=1,2,3$) are the Pauli matrices. 
These satisfy the Clifford algebra 
\be
 \brc{\gm^\mu,\gm^\nu}=2\eta^{\mu\nu}. 
\ee
The $\SUu$ Majorana condition is 
\be
 \bar{\Omg}^{Ii}\equiv(\Omg^I_i)^\dagger\gm_0=(\Omg^{Ii})^T C_5, 
 \label{su_mjrn}
\ee
where the 5D charge conjugation matrix~$C_5$ is 
\be
 C_5=\mtrx{-i\sgm_2}{}{}{i\sgm_2}. 
\ee

The chirality matrix~$\gm_5$ in four dimensions is defined as 
$\gm_5\equiv i\gm^4$, and  
the projection operators~$\cP_{\rm R,L}$ are defined as 
\be
 \cP_{\rm R} \equiv \frac{1+\gm_5}{2}. \;\;\;\;\;
 \cP_{\rm L} \equiv \frac{1-\gm_5}{2}.  \label{prj_op}
\ee
In the superfield formalism, we use the 2-component notation for spinors. 
For a 4-component spinor~$\psi$, a 2-component spinor~$\psi_{\rm R}$ is 
defined as 
\be
 \cP_{\rm R}\psi \equiv \vct{\psi_{\rm R}}{0}. \label{def_RH}
\ee
Using this notation, we can rewrite the 4-component spinors~$\Omg^{Ii}$ 
in terms of 2-component spinors~$\Omg^{Ii}_{\rm R}$ 
owing to the $\SUu$ Majorana condition~(\ref{su_mjrn}). 
\be
 \Omg^{I1}=\vct{\Omg^{I1}_{\rm R}}{-\bar{\Omg}^{I2}_{\rm R}}, \;\;\;\;\;
 \Omg^{I2}=\vct{\Omg^{I2}_{\rm R}}{\bar{\Omg}^{I1}_{\rm R}}. 
\ee
Similarly, we can rewrite the $\Usp$ Majorana spinors~$\zeta^\alp$ as, 
\be
 \zeta^1=\vct{\ztR^1}{-\bar{\zeta}^2_{\rm R}}, \;\;\;
 \zeta^2=\vct{\ztR^2}{\bar{\zeta}^1_{\rm R}}, \;\;\;
 \zeta^3=\vct{\ztR^3}{-\bar{\zeta}^4_{\rm R}}, \;\;\;
 \zeta^4=\vct{\ztR^4}{\bar{\zeta}^3_{\rm R}}, \;\;\; \cdots. 
\ee 
We take a convention of Ref.~\cite{WB} 
for the contraction of 2-component spinors. 

The indices~$\alp,\bt,\cdots$ and $i,j,\cdots$ 
are lowered (or raised) by the antisymmetric tensors~$\rho_{\alp\bt}$ 
($\rho^{\alp\bt}$) and $\ep_{ij}$ ($\ep^{ij}$), 
respectively.\footnote{
We take the northwest-to-southeast contraction convention 
for these indices. }
Here, $\ep_{12}=\ep^{12}=1$ and both $\rho_{\alp\bt}$ and $\rho^{\alp\bt}$ 
have the following form in the standard representation \cite{dewit}.  
\be
 \rho=\ep\otimes\id_{\nH+1}. \;\;\;\;\;
 (\ep=i\sgm_2)
\ee

\section{Invariant action on gravitational background} \label{inv_act}
After performing the integration of $\tht$ and 
substituting Eqs.(\ref{sf_comp1}) and (\ref{sf_comp2}) 
into each component of superfield, 
the superspace expression~(\ref{Sinv}) becomes 
\bea
 S \eql \int\dr^5x\;\cL, \nonumber\\
 \cL \eql e\left[-\frac{\cN_{IJ}}{2}\brkt{
 -\frac{1}{4}\cF^{I\mu\nu}\cF_{\mu\nu}^J+\frac{1}{2}\der^\mu M^I\der_\mu M^J
 -Y^{Ii}_{\;\;\;\;j}Y^{Jj}_{\;\;\;\;i}
 +2i\bar{\Omg}^{Ii}\gm^\mu\cD_\mu\Omg^J_i} 
 \right. \nonumber\\
 &&\hspace{5mm}\left. +\cN_{IJK}\brkt{\frac{i}{4}\bar{\Omg}^{Ii}\gm^{\mu\nu}
 \Omg^J_i F^K_{\mu\nu}+i\bar{\Omg}^{Ii}Y^J_{ij}\Omg^{Kj}}\right]
 +\frac{1}{8}C_{IJK}\ep^{\mu\nu\rho\sgm\tau}W_\mu^IF_{\nu\rho}^JF_{\sgm\tau}^K
 \nonumber\\
 &&+e\left[\brkt{1+\frac{(\gey^{-1}W_y^0)^2}{(M^0)^2}}\tl{\cF}^\alp_i\dmx
 \tl{\cF}^i_\bt+\cD^\mu\cA_i^\alp\dmx
 \cD_\mu\cA_\bt^i-2i\bar{\zeta}^\alp\dmx\gm^\mu\cD_\mu\zeta_\bt 
 \right.\nonumber\\
 &&\hspace{8mm}+\cA_i^\alp\dmx\brkt{M^I(gt_I)M^J(gt_J)}_\bt^{\;\;\gm}
 \cA_\gm^i-8i\cA_i^\alp\dmx\bar{\Omg}^{Ii}(gt_I)_\bt^{\;\;\gm}\zeta_\gm
 \nonumber\\
 &&\hspace{8mm}\left.+2\cA_i^\alp\dmx Y^{Iij}(gt_I)_\bt^{\;\;\gm}\cA_{\gm j}
 +2i\bar{\zeta}^\alp\dmx M^I(gt_I)_\bt^{\;\;\gm}\zeta_\gm\right] \nonumber\\
 &&-e\brkt{\frac{1}{8}\cN-\frac{3}{16}\cA^\alp_i\dmx\cA^i_\bt}\cR, 
 \label{L_comp}
\eea
where summations for the indices~$I,J,K=0,1,\cdots,\nV$; 
$\alp,\bt,\gm=1,2,\cdots,2\nH+2$, and $i,j=1,2$ are implicit, and 
$\cN_{IJ}\equiv\der^2\cN/\der M^I\der M^J$, 
$\cN_{IJK}\equiv\der^3\cN/\der M^I\der M^J\der M^K$. 
Here, we have used the 5D spinor notation. 
Namely, $\bar{\Omg}^{Ii}$ ($\bar{\zeta}^\alp$) denote 
$\SUu$ ($\Usp$) Majorana conjugates of $\Omg^{Ii}$ ($\zeta^\alp$).  
$\cR$ is the background value of 5D Ricci scalar and  
\bea
 e \defa \det(\ge{\mu}{\udl{\nu}})=e^{4\sgm}, \nonumber\\
 \tl{\cF}_i^\alp \defa \cF_i^\alp-M^0(gt_0)^\alp_{\;\;\bt}\cA_i^\bt, 
 \nonumber\\
 \cD_\mu\cA_i^\alp \defa \der_\mu\cA^\alp_i-V_{\mu ij}\cA^{\alp j}
 -W^I_\mu(gt_I)^\alp_{\;\;\bt}\cA_i^\bt, \nonumber\\
 \cD_\mu\zeta^\alp \defa \brkt{\der_\mu-\frac{1}{4}\gomg{\mu}
 {\udl{\nu}\udl{\rho}}\gm_{\udl{\nu}\udl{\rho}}}\zeta^\alp
 -W^I_\mu(gt_I)^\alp_{\;\;\bt}\zeta^\bt, \nonumber\\
 \cR \eql 4\gey^{-2}(2\ddot{\sgm}+5\dot{\sgm}^2). 
\eea
Eq.(\ref{L_comp}) coincides with the invariant action 
in Refs.~\cite{KO1,Ohashi} 
on the gravitational background~(\ref{grv_bgd}), 
except for the following three points. 
First, the coefficient of $\tl{\cF}_i^\alp\dmx\tl{\cF}_\bt^i$ is 
$1-W^{0\mu}W^0_\mu/(M^0)^2$ in Refs.~\cite{KO1,Ohashi}. 
However, this discrepancy is harmless because $W^0_m$ is odd 
under the orbifold parity and vanishes on the boundary. 
Both actions lead to the same on-shell action.\footnote{
The linear term for $\tl{\cF}_i^\alp$ only comes from the brane-localized 
actions \cite{KO2}. }
Second, the coefficient of the Einstein-Hilbert term is different 
from that of Ref.~\cite{KO1,Ohashi}, which is $e\cA_i^\alp\dmx\cA_\bt^i/4$. 
This discrepancy stems from the fact that we have dropped 
the auxiliary field~$D$ that leads to the constraint~(\ref{ctrt1}). 
In fact, both coefficients will be the correct value~$-eM_5^3/2$ 
after the gauge fixing. 
Third, according to Refs.~\cite{KO1,Ohashi}, the coefficient functions of 
the kinetic term for the gauge fields and gauge scalars 
are $\cN a_{IJ}$ defined in Eq.(\ref{geo_qnt}) 
instead of $-\cN_{IJ}/2$ in Eq.(\ref{L_comp}). 
This stems from the fact that 
we have replaced the auxiliary field~$v_{\udl{\mu}\udl{\nu}}$ 
in the Weyl multiplet by its background value. 
However, this does not cause a serious problem if we do not discuss 
the graviphoton sector. 
In fact, after the gauge fixing, 
the difference from the correct one only affects 
in the quartic or higher order terms that are neglected here, 
except for the graviphoton sector. 
Therefore, Eq.(\ref{Sinv}) effectively reproduces the invariant action 
in Refs.~\cite{KO1,Ohashi}. 

From Eq.(\ref{L_comp}), we can obtain the on-shell expressions of 
$V_y^{(r)}$ and $Y^{I=0(r)}$ ($r=1,2,3$).
\bea
 V_y^{(3)} \eql -\frac{1}{M_5^3}\Im\brkt{\bar{\cA}_2^3\der_y\cA_2^3
 +\bar{\cA}_2^4\der_y\cA_2^4}+\cdots, \nonumber\\
 V_y^{(1)}+iV_y^{(2)} \eql -\frac{i}{M_5^3}\brkt{
 \bar{\cA}_2^3\der_y\bar{\cA}_2^4-\bar{\cA}_2^4\der_y\bar{\cA}_2^3}+\cdots, 
 \nonumber\\
 Y^{0(3)} \eql -\frac{M_5^2 g_{\rm c}^0}{3}\brkt{1+\frac{
 \abs{\cA_2^3}^2+\abs{\cA_2^4}^2}{M_5^3}}-\frac{g_{\rm h}^0}{3M_5}\brkt{
 \abs{\cA_2^3}^2-\abs{\cA_2^4}^2}+\cdots, \nonumber\\
 Y^{0(1)}+iY^{0(2)} \eql 
 \frac{2g_{\rm h}^0}{3M_5}\bar{\cA}_2^3\bar{\cA}_2^4+\cdots, 
 \label{Y12}
\eea
where the ellipses denote terms that have vanishing VEVs. 
We have used the gauge fixing conditions. 
$Y^{I\neq 0(r)}$ does not involve the stabilizer fields~$\cA_2^{\alp=3,4}$, 
and have vanishing VEVs under the assumption of Eq.(\ref{gauge_cp}).

\end{document}